\documentclass[letterpaper]{jpconf}
\usepackage{graphicx}

\newcommand{\be}{\begin{equation}}
\newcommand{\ee}{\end{equation}}
\newcommand{\n}[1]{\label{#1}}
\newcommand{\eq}[1]{(\ref{#1})}
\newcommand{\hhh}{\, ,\hspace{0.2cm}}

\begin{document}
\title{Charged black strings in 5D Kaluza-Klein spacetime}

\author{Andrey A. Shoom}

\address{Theoretical Physics Institute, University of Alberta, Edmonton, AB, Canada,  T6G 2G7}

\ead{ashoom@phys.ualberta.ca}

\begin{abstract}
We present a short review on the electric and magnetic black strings in 5D spacetime with one compact dimension.  Linear static perturbations of these objects indicate presence of the threshold unstable mode. Analysis of the mode shows that the electric black string is less stable than a neutral one. The situation is opposite for the magnetic black string.            
\end{abstract}

\section{Introduction}

In attempts to resolve the hierarchy problem existence of large extra spatial dimensions was proposed in the modern theoretical models \cite{ADD1}, \cite{ADD2}. Existence of large compact extra dimensions opens an interesting possibility of mini black holes production in the future LHC experiments (see e.g. \cite{DL}, \cite{YN}, \cite{G}). In addition, in spacetimes with large compact extra dimensions a large variety of topologically different black objects may exists: black holes, black rings, black branes, black strings, etc. (see e.g. \cite{HaNOb}). Study of such objects, especially topological phase transitions between them, is a subject of high interest.

In this paper we study static electric and magnetic black strings in 5D spacetime with one large compact dimension. Horizon surface topology of such black strings is $S^2\times S^1$, which is different from the spherical topology $S^3$ of a 5D {\it localized black hole}. Such black strings and black holes are considered to be different topological phases in a topological phase transition diagram. Topological phase transitions are closely related to thermodynamic phase transitions of black objects, which are thermodynamic systems. Black objects possess entropy (the Bekenstein-Hawking entropy which is proportional to a quarter of horizon surface area), temperature, and other thermodynamic characteristics. Entropy of a neutral 5D black string of mass $M$ is $S_{BS}\propto M^2$, whereas entropy of a 5D localized black hole of the same mass is approximated\footnote{Exact solution representing such black hole is not known. Analytical approximations are given in \cite{GK1}, \cite{GK2}, \cite{KS}.} by $S_{BH}\propto M^{3/2}L^{1/2}$. Here, $L$ is the asymptotic size of the compact dimension. Thus, if $M>L$, up to a constant factor, the black string entropy is larger than the black hole entropy. Hence, the black string topological phase is globally thermodynamically stable. However, for a larger size of the compact dimension, $S_{BH}$ can be greater than $S_{BS}$. Thus, as it was suggested in \cite{GL}, the black string may be unstable under large wavelength gravitational perturbations. Such instability (the so-called Gregory-Laflamme instability) is a generic property of many black objects. It was studied for a dilaton black string with magnetic charge \cite{GL2}, for boosted black strings \cite{Myers}, for magnetic black strings \cite{Miya}, and for electric black strings in connection with bubble spacetimes \cite{SL}. Topological transition between black strings in 5D and 6D spacetimes, and the corresponding localized black holes was studied in \cite{Jutta}. For detailed reviews on the phase transitions between black strings and black holes see e.g. \cite{HaNOb}, \cite{Kol}.

In this paper we present a short review on the Gregory-Laflamme instability of electric and magnetic 5D black strings based on results published in \cite{Miya}, \cite{SL}, \cite{FS}. In section 2 we present the black string metrics and discuss their properties. Section 3 contains a review of numerical results of linear, static, gravitational perturbations of the strings and the topological phase transition diagrams. In section 4 we summarize our results.
  
\section{5D charged black strings solutions}

Let us consider the following action\footnote{We shall use the following units convention: $16\pi G_{(4)}=c=\hbar=k_B=1$.}
\be\n{S}
S=\frac{1}{16\pi G_{(5)}}\int_0^L dz\int dx^4\sqrt{-g}\left(R-\frac{1}{4}F_{\alpha\beta}F^{\alpha\beta}\right)\hhh F_{\alpha\beta}=\nabla_{\alpha}A_{\beta}-\nabla_{\beta}A_{\alpha}.
\ee
Here, $G_{(5)}$ is the 5D gravitational constant (5D and 4D gravitational constants are related as follows: $G_{(5)}=G_{(4)}L$), $R$ is the 5D Ricci scalar, $A_{\alpha}$ is the 5D electromagnetic vector potential. Here, and in what follows $\nabla_{\alpha}$ denotes a covariant derivative defined with respect to the 5D metric $g_{\alpha\beta}$. The corresponding Einstein-Maxwell equations obtained from the principle of least action are: 
\be\n{EE}
R_{\alpha\beta}-\frac{1}{2}g_{\alpha\beta}R=\frac{1}{2}T^{(em)}_{\alpha\beta}\hhh T^{(em)}_{\alpha\beta}=F_{\alpha}^{\,\,\,\gamma}F_{\beta\gamma}-\frac{1}{4}g_{\alpha\beta}F_{\gamma\delta}F^{\gamma\delta}\hhh \nabla_{\beta}F^{\alpha\beta}=0\hhh \nabla_{[\gamma}F_{\alpha\beta]}=0.
\ee
The mass $M$ of a 5D black object is defined by Komar's formula \cite{MP}
\be\n{M}
M=\frac{1}{16\pi G_{(5)}}\frac{3}{2}\oint_{\Sigma_{(3)}^{\infty}}\frac{dx^{\gamma}dx^{\delta}dx^{\lambda}}{\sqrt{-g}}\varepsilon_{|\gamma\delta\lambda|\alpha\beta}\nabla^{\alpha}\xi^{\beta},
\ee
where $\xi^{\alpha}$ is the timelike Killing vector normalized at infinity as $\xi^{\alpha}\xi_{\alpha}=-1$. Here, $\varepsilon^{01234}=+1$, $\varepsilon_{01234}=g$, and $|\alpha\beta\cdots\gamma|=\alpha<\beta<\cdots<\gamma$ means proper orientation. The electric charge $Q$, and the magnetic charge $P$ are defined as follows:
\be\n{Q}
Q=-\frac{1}{2}\oint_{\Sigma_{(3)}}\frac{dx^{\gamma}dx^{\delta}dx^{\lambda}}{\sqrt{-g}}\varepsilon_{|\gamma\delta\lambda|\alpha\beta}F^{\alpha\beta}\hhh P=-\oint_{\Sigma_{(2)}}dx^{\alpha}dx^{\beta}F_{|\alpha\beta|}.
\ee

\subsection{Electric black string}

Solution representing static, uniform electric black string is given by \cite{SL}, \cite{Jutta}, \cite{FS}:
\be\n{EBS}
ds^2=-\frac{r(r-w)}{(r+h)^2}dt^2+\frac{r+h}{r-w}dr^2+\frac{r+h}{r}dz^2+r(r+h)d\omega^2\hhh A_{\alpha}=-\frac{\sqrt{3h(w+h)}}{r+h}\delta_{\alpha}^{\,\,\,t},
\ee
where $d\omega^2=d\theta^2+\sin^2\theta d\phi^2$, is the metric on a unit 2D round sphere, and the coordinate ranges are: $t\in(-\infty,\infty)\hhh z\in[0,L)\hhh r\in[0,\infty)\hhh \theta\in[0,\pi]\hhh \phi\in[0,2\pi)$. We denote $(t,r,z,\theta,\phi)=(x^0,x^1,x^2,x^3,x^4)$. Using the definitions of mass \eq{M} and electric charge \eq{Q} we derive:
\be\n{Me}
M=6\pi(w+2h)\hhh Q=4\pi L\sqrt{3h(w+h)}\hhh w=\frac{1}{6\pi}\sqrt{M^2-3(Q/L)^2}\hhh h=\frac{M}{12\pi}-\frac{w}{2}.
\ee
Thus, the electric charge is defined within the range: $0\leq|Q|\leq ML/\sqrt{3}$. Horizon of the black string is defined by $r=w\geq0$. The spacetime \eq{EBS} is singular at $r=0$. Thus, the extremal black string horizon $r=w=0$ is singular. Calculating the energy-momentum tensor components in a local tetrad frame we derive the energy density $\rho$, and the principal pressures $p_i$, $(i=r,z,\theta,\phi)$: $\rho=-p_r=p_z=p_{\theta}=p_{\phi}=3h(w+h)[2r(r+h)^3]^{-1}$. Thus, all but $p_r$ principal pressures are positive.

\subsection{Magnetic black string}

Solution representing static, uniform magnetic black string is given by \cite{Miya}, \cite{FS}:
\be\n{MBS}
ds^2=-\frac{r-w}{r+h}dt^2+\frac{(r+h)^2}{r(r-w)}dr^2+\frac{r}{r+h}dz^2+(r+h)^2d\omega^2\hhh A_{\alpha}=\sqrt{3h(w+h)}\cos\theta\delta_{\alpha}^{\,\,\,\phi}.
\ee
Using the definitions of mass \eq{M} and magnetic charge \eq{Q} we derive:
\be\n{Mm}
M=6\pi(w+h)\hhh P=4\pi\sqrt{3h(w+h)}\hhh w=\frac{1}{6\pi M}(M^2-3/4P^2)\hhh h=\frac{P^2}{8\pi M}.
\ee
The spacetime is singular at $r=-h<0$, and can be analytically continued through $r=0$. The black string horizon is defined by $r=w>-h$. To preserve the spacetime signature we define $w\geq0$ corresponding to $0\leq|P|\leq 2M/\sqrt{3}$. The energy-momentum tensor components calculated in a local tetrad frame give: $\rho=-p_r=p_{\theta}=p_{\phi}=-p_z=3h(w+h)[2(r+h)^4]^{-1}$. Thus, in contrast to the electric black string $p_z$ principal pressure is negative.

\section{Static perturbations of the black strings}

We consider spherically symmetric gravitational perturbations of the black string spacetimes: $g_{\alpha\beta}\to g_{\alpha\beta}+h_{\alpha\beta}$, where $h_{\alpha\beta}\ll 1$. Fourier mode of a spherically symmetric wave propagating in $z$ direction is $h_{\alpha\beta}=Re(a_{\mu\nu}(r)e^{-i\omega t+ikz})$. It is unstable if $\omega=i\Omega$, and $\Omega>0$. Time-independent (critical) mode $\Omega=0$ defines the threshold mode of the Gregory-Laflamme instability. To find such mode we consider spherically symmetric, static perturbations: $h_{\alpha\beta}=\{h_{tt},h_{rr},h_{rz},h_{\theta\theta},h_{\phi\phi}=h_{\theta\theta}\sin^2\theta, h_{zz}\}$.
Analysis of the Einstein-Maxwell equations for the first order perturbations shows that induced electromagnetic perturbations decouple from the gravitational ones and can be neglected (for details see \cite{Miya}, \cite{SL}, \cite{FS}). According to the numerical results the instability starts at the special value of the wave number $k=k_{cr}$ corresponding to the minimal wavelength $\lambda_{cr}=2\pi/k_{cr}$. Modes with smaller values of wavelength are stable. Such behavior is analogous to the Jeans instability. If the asymptotic size of the extra compact dimension, $L$, is smaller than the minimal wave-length, $\lambda_{cr}$, then unstable modes can not be accommodated into the compact dimension, and the instability does not set in. However, if $L\geq\lambda_{cr}$, the black strings are unstable. Numerical analysis shows that for the electric black string the dimensionless wave number $\kappa_{cr}:=wk_{cr}\approx0.876$ does not depend on the electric charge value, and equals to that of a neutral 5D black string. Setting $L=\lambda_{cr}$ and defining the dimensionless mass $\mu=M/L$ and electric charge $q=Q/L^2$ we derive with the aid of \eq{Me} the following relation
\be\n{el}
\mu=\sqrt{9\kappa_{cr}^2+3q^2}.
\ee
An analogical relation can be derived using the global thermodynamic stability argument. Assuming that another topological phase of the black string is a 5D electric localized black hole, we approximate the black hole entropy with the entropy of a 5D Reissner-Nordstr\"om black hole \cite{MP}, and derive the micro-canonical critical curve $S_{BS}=S_{BH}$ (for details see \cite{FS})
\be\n{elt}
\mu=\sqrt{9/4+3q^2}.
\ee
In the case of the magnetic black string $\kappa_{cr}$ depends on the value of the magnetic charge. Defining the dimensionless magnetic charge $p=P/L$ and using \eq{Mm} we derive
\be\n{mag}
\mu=\frac{3\kappa_{cr}(m)}{1-m}\hhh p=\frac{2\kappa_{cr}(m)\sqrt{3m}}{1-m}\hhh m=\frac{h}{h+w}.
\ee
The relations \eq{el}, \eq{elt} and \eq{mag} are illustrated in Figures 1 and 2. From Figure 1 we observe thermodynamic and dynamic correspondence for the electric black string. One may expect a decay of the unstable electric black string into the corresponding localized black hole. However, the question if such transition is possible remains open \cite{HM}. The situation is different for the magnetic black string. The magnetic charge $P$ is related to $2$-form $F_{\alpha\beta}$ (see \eq{Q}), whereas magnetic charge of a 5D localized black hole, which is dual to the electrically charged one, is related to $3$-form $F_{\alpha\beta\gamma}$ which does not belong to our theory \eq{S}. Thus, the final state of the magnetic black string topological phase transition is not known.

\begin{figure}[h]
\begin{minipage}{8.5cm}\hspace{2.0cm}%
\includegraphics[height=4.5cm,width=4.48cm]{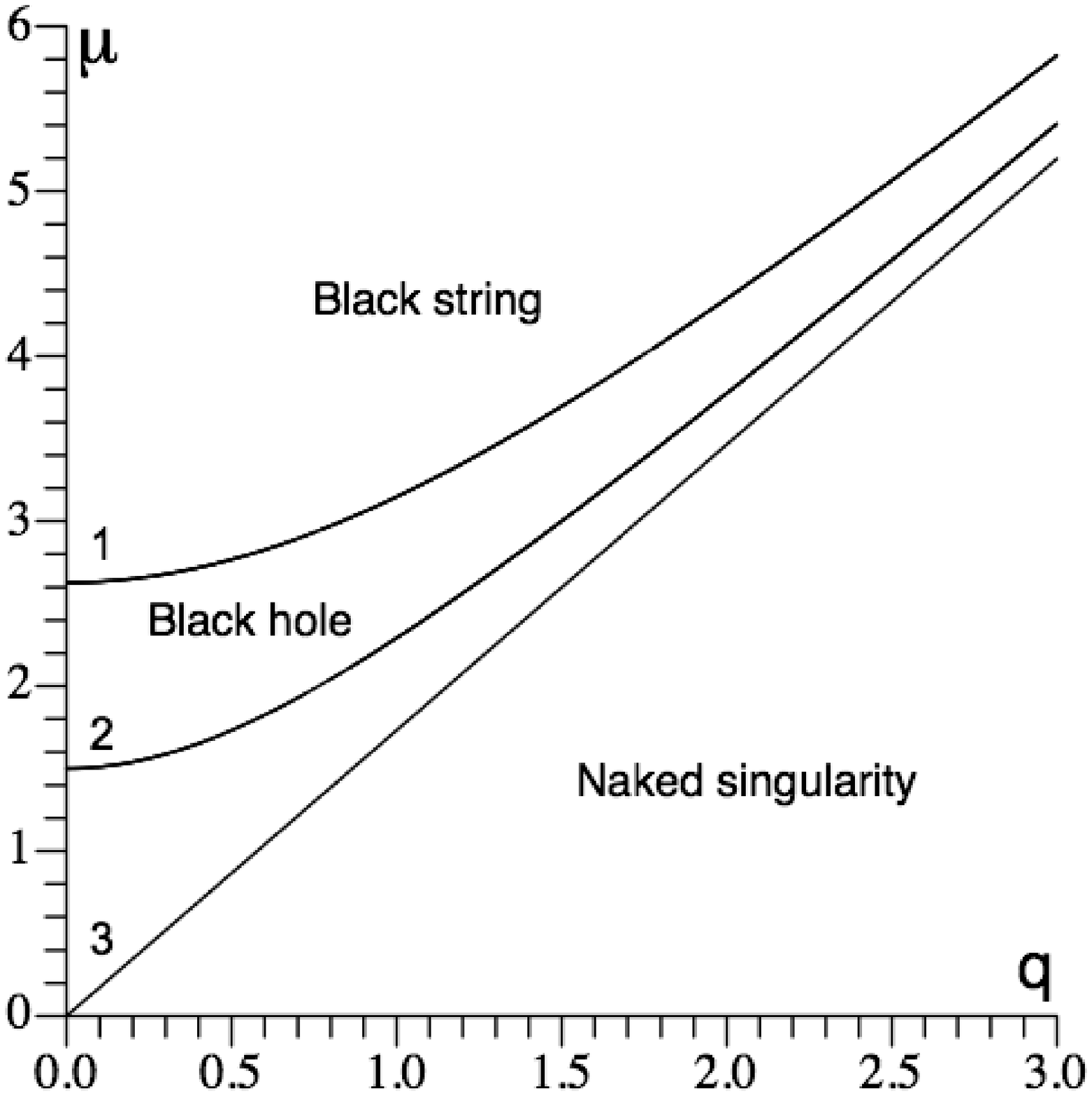}
\caption{\label{f1} Curve (1) and curve (2) illustrate the relations \eq{el} and \eq{elt}, respectively. They separate topological phases of the electric black string and the corresponding localized black hole. Line (3) defines the extremal black string: $\mu=q\sqrt{3}$.}
\end{minipage}\hspace{0.5cm}%
\begin{minipage}{6.99cm}\hspace{0cm}%
\includegraphics[height=4.5cm,width=6.99cm]{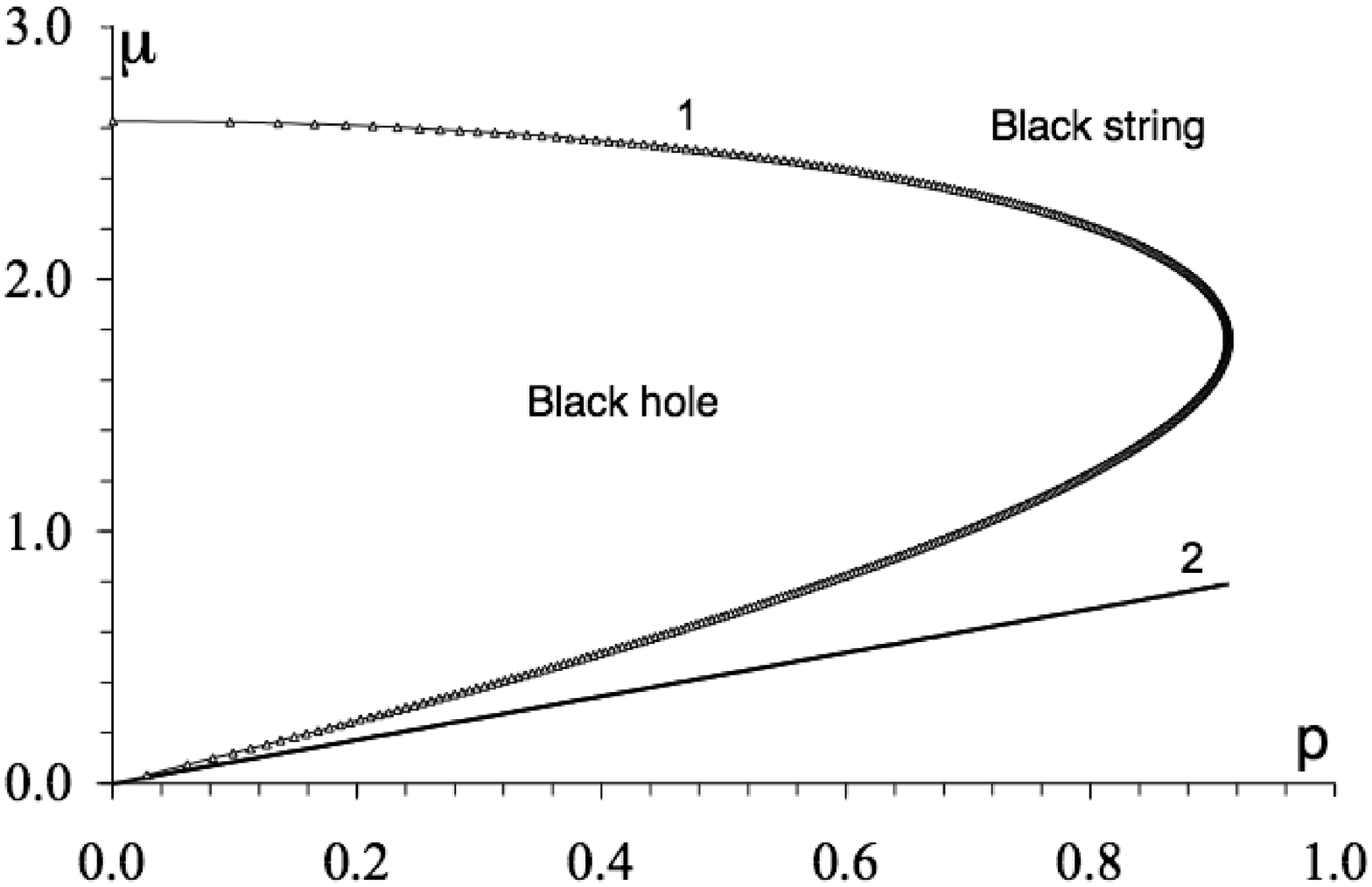}
\caption{\label{f2} The relation \eq{mag} is illustrated by curve (1), which separates topological phases of the magnetic black string and the corresponding localized black hole. Line (2) corresponds to $p=2\mu/\sqrt{3}$.}
\end{minipage} 
\end{figure} 
\vspace{-0.5cm}  

\section{Summary}

Our results show that the electric (magnetic) black string is less (more) stable than a neutral one. Thus, the electric charge tends to destabilize black string, whereas the magnetic charge makes it more stable. This can be deduced from the form of $g_{zz}$ components of the corresponding metrics. For the electric (magnetic) black string, near the horizon $g_{zz}>1$ ($g_{zz}<1$), that makes the proper length along the compact dimension greater (smaller) than $L$. Hence, the electric (magnetic) black string is less (more) stable. This result is related to the fact that the principal pressure $p_z$ is positive for the electric black string and negative for the magnetic black string.


\section{References}
\begin{thebibliography}{}

\bibitem{ADD1} Arkani-Hamed N, Dimopoulos S, Dvali G 1998 {\it Phys. Lett.} B {\bf 429} 263
\bibitem{ADD2} Antoniadis I, Arkani-Hamed N, Dimopoulos S, Dvali G 1998 {\it Phys. Lett.} B {\bf 436} 257
\bibitem{DL} Dimopoulos S, Landsberg G 2001 {\it Phys. Rev. Lett.} {\bf 87} 161602
\bibitem{YN} Yoshino H and Nambu Y 2003 {\it Phys. Rev.} D {\bf 67} 024009
\bibitem{G} Gingrich D M 2006 {\it Int. J. Mod. Phys} A {\bf 21} 6653
\bibitem{HaNOb} Harmark T, Niarchos V and Obers N A 2007 {\it Class. Quantum Grav.} {\bf 24} R1
\bibitem{GK1} Gorbonos D and Kol B 2004 {\it J. High Energy Phys.} JHEP06(2004)053
\bibitem{GK2} Gorbonos D and Kol B 2005 {\it Class. Quantum Grav.} {\bf 22} 3935
\bibitem{KS} Karasik D, Sahabandu C, Suranyi P and Wijewardhana L C R 2005 {\it Phys. Rev.} D {\bf 71} 024024
\bibitem{GL} Gregory R and Laflamme R 1993 {\it Phys. Rev. Lett.} {\bf 70} 2837
\bibitem{GL2} Gregory R and Laflamme R 1994 {\it Nucl. Phys.} B {\bf 428} 399
\bibitem{Myers} Hovdebo J L and Myers R C 2006 {\it Phys. Rev.} D {\bf 73} 084013
\bibitem{Miya} Miyamoto U 2008 {\it Phys. Lett.} B {\bf 659} 380
\bibitem{SL} Sarbach O, Lehner L 2005 {\it Phys. Rev.} D {\bf 71} 026002
\bibitem{Jutta} Kleihaus B, Kunz J and Radu E 2006 {\it J. High Energy Phys.} JHEP06(2006)016
\bibitem{Kol} Kol B 2006 {\it Phys. Rep.} {\bf 422} 119
\bibitem{FS} Frolov V P and Shoom A A 2009 {\it Phys. Rev.} D {\bf 79} 104002
\bibitem{MP} Myers R C and Perry M J 1986 {\it Annals of Physics}  {\bf 172} 304
\bibitem{HM} Horowitz G T and Maeda K 2001 {\it Phys. Rev. Lett.} {\bf 87} 131301

\end{thebibliography}
\end{document}